\documentclass[aps,pra,twocolumn,showpacs,floatfix]{revtex4}
\usepackage[utf8x]{inputenc}
\usepackage{graphicx}
\usepackage{epsfig}
\usepackage{color}
\usepackage{amsmath}
\usepackage{amssymb}
\usepackage{dsfont}
\usepackage{hyperref}

\newcommand{\hn}{{\hat n}}

\newcommand{\br}{ \mathbf{r} } 
 
\newcommand{\bq}{ \mathbf{q} }
\newcommand{\bk}{ \mathbf{k} }
\newcommand{\bb}{ \mathbf{b} }

\newcommand{\CO}{{(Color online.)\;}}

\begin{document}

\title{
Absence of quasiclassical coherence in mean-field dynamics of bosons in a kinetically frustrated regime
}
\author{Akos Rapp}
\affiliation{ Institut f\"ur Theoretische Physik, Leibniz Universit\"at, 30167 Hannover, Germany }
\date{\today}

\begin{abstract}
We study numerically the dynamics of bosons on a triangular lattice after quenching both the on-site interactions and the external trapping potential
to negative values. In a similar situation on the square lattice, the dynamics can be understood in terms of an effectively reversed Hamiltonian.
On the triangular lattice, however, the kinetic part of the reversed Hamiltonian is frustrated and whether coherence can develop is an open question. 
The strength of the frustration can be changed by tuning the ratio of the hopping rates along different directions.
We calculate time-of-flight images at different times after the quench for different values of the hopping anisotropy. 
We observe peaks at the maxima of the noninteracting dispersion relation both in the isotropic case
and also in the rhombic limit of high hopping anisotropy showing quasiclassical coherence.
For an intermediate value, however, no coherence develops up to the longest simulation times.
These results imply that experiments along similar lines could study unconventional superfluidity of bosons and aspects
of the conjectured spin-liquid behavior in the hard-core limit.
\end{abstract}


\maketitle

\section{Introduction}

A broadly observed fact is that most physical systems undergo
transitions to phases of matter which display some kind of order
as the system is cooled down. In geometrically frustrated 
systems, however, such ordering may not be possible down to the
lowest temperatures since the spatial arrangement is incompatible
with certain order types.
For classical Ising spins on the triangular lattice
with antiferromagnetic coupling, this problem was first discussed by Wannier~\cite{Wannier},
who found finite ground-state entropy, a consequence of a highly degenerate ground state.
In three dimensions, frustration plays an important role in
spin ice materials, which display magnetic monopoles~\cite{spinice}, 
the microscopic version of the hypothetical cosmic counterpart proposed in the famous paper 
by Dirac~\cite{Dirac}.
Strong quantum fluctuations present for lower spin lengths can give rise to
elusive spin-liquid phases~\cite{Balents}.

The identification of spin liquids in solid state systems is very challenging,
on one hand, due to the featureless nature of the spin-liquid phase, but also
because of the interplay of additional degrees of freedom, phonons, disorder, etc.
In contrast, ultracold atoms in optical lattices (see the review in Ref.~\cite{Blochreview}) present
exceptionally clean systems where microscopic parameters can be tuned experimentally
in a broad range with great control. In addition, the relatively large typical spatial and time scales 
allow for tracking physical processes more easily. Ultracold atoms in optical
lattices are therefore ideal quantum simulators for many-body systems.

Fermionic atoms in a lattice with two spin components
at half-filling with strong on-site repulsion  are described
by the antiferromagnetic Heisenberg model at low energies. This implies
that on a triangular lattice, such fermions could naturally exhibit frustrated quantum
magnetism. However, the entropies of fermionic clouds are relatively high~\cite{NeelAFM}, and
spontaneous antiferromagnetic long-range order has not yet been established
even on bipartite lattices. The reason behind this is well known: due to Pauli blocking,
fermionic clouds cannot be cooled efficiently while the typical temperature to be reached is
relatively low, of the order of the Heisenberg exchange energy for the Mott insulator.
We will concentrate on bosonic atoms in optical lattices in the following.

While frustration originates from the interaction terms of the Hamiltonian in 
triangular magnets or fermionic Mott insulators, in spinless bosonic systems 
it is related to the kinetic energy. Nevertheless, the realization of kinetic
frustration is not straightforward in optical lattices.
One reason behind this is that the nearest-neighbor hopping amplitude $J$
between lattice sites in the lowest Bloch band has a definite sign, usually defined
with the convention $J>0$. 
This is implied physically by the fact that the lowest energy usually implies zero momentum 
and mathematically by the solution of the Mathieu equation describing 
the one-body problem in the one-dimensional standing-wave optical lattice potential.
Simple square or cubic lattices built from this potential naturally share this property;
moreover, the non-interacting diagonal (next-nearest-neighbor) hopping is 
exactly zero due to the separability of the one-body problem. Therefore, 
optical lattice setups need to overcome two problems to realize frustration:
the lack of triangular graphs and the sign of the hopping. The former can be
solved by using a different lattice geometry.

While certain hopping amplitudes in higher Bloch bands have opposite signs,  
such systems are not especially suitable for quantum simulations. One issue is the high instability with respect to decay 
to other bands, which is not easily circumvented~\cite{multibandBloch,multibandHemmerich}.

A very successful idea to effectively change the sign of the hopping amplitude $J \to J_{\rm eff}$ is based
on a periodic shaking of the optical lattice~\cite{shaking0}. This idea lead to
a proposal for bosons on the triangular lattice with an elliptical lattice 
shaking~\cite{shaking-EPL}, which allows for a continuous tuning of the effective 
hopping anisotropically. Bosons can be mapped to XY spin models in two limits
of the interaction strength. For weak interactions and sufficiently high
boson filling, each site can be described as an individual ``superfluid'' droplet
with a well-defined phase and the bosonic Hamiltonian can be mapped to a \emph{classical} XY model~\cite{FisherFisher}. 
In the other limit, at half-filling and infinitely strong repulsion,
the Hamiltonian can be mapped to a \emph{quantum} XY model~\cite{shaking-EPL}.
Both of these models on a triangular lattice are 
frustrated with antiferromagnetic couplings, given
by a negative effective hopping, $J_{\rm eff} < 0$.
In the case of isotropic nearest-neighbor spin couplings, it is believed
that both ground states exhibit 120$^\circ$ spiral 
long-range magnetic order
[U(1) rotational symmetry breaking],
with a non-zero chirality ($Z_2$ symmetry breaking)
 ~\cite{classical-triang-XY,quantum-triang-XY}. 
As the anisotropy of the couplings is increased, the chirality
decreases and vanishes. In the classical model this happens
at the so-called rhombic transition point, beyond which
only the U(1) spin symmetry is broken.
Most interestingly, it was proposed that in the hard-core limit, instead of 
a single phase transition, a gapped spin-liquid phase 
emerges between two quasi-classically ordered phases~\cite{quantum-triang-XY,scatl}.
The conjectured phase diagram for the bosons is displayed in Ref.~\cite{shaking-EPL}.
While the lattice shaking technique succeeded experimentally in simulating 
frustrated classical magnetism~\cite{shaking-exp,shaking-exp2},
no signature of the quantum magnetism has been reported so far.
To study the quantum aspects of frustration 
with ultracold bosons, a new approach is required.

An alternative route to reversed hopping in the lowest Bloch band
is employing negative absolute temperatures, $T<0$. 
Negative absolute temperatures can be reached in closed systems with Hamiltonians with an upper bound~\cite{Landau}. 
With ultracold atoms in optical lattices, such a Hamiltonian $H$ can be engineered
basically by switching the sign of the external harmonic trapping potential $V_0 > 0 \to V_0 < 0$~\cite{Mosk,negT,negT-sim,negT-exp,negT-tdGA,negT-1D}. 
Energy conservation restricts the dynamics and the atomic cloud cannot explode as long as the kinetic energy
is bounded. This important condition is provided by a sufficiently deep optical lattice.
In equilibrium at $T<0$, the partition function of the system is equivalent to the partition
function of a system at an effective temperature $|T|$ governed by the reversed Hamiltonian $-H$. 
In Ref.~\cite{negT-sim} it was discussed that this mapping can be used to 
simulate Hamiltonians that have couplings with signs that are hard to reach experimentally. 
In this work we apply this idea to bosons on the triangular lattice, which, 
only at negative $T$, have ``frustrated'' kinetic energy.

\begin{figure}
  \centering
  \includegraphics[width=0.45\textwidth]{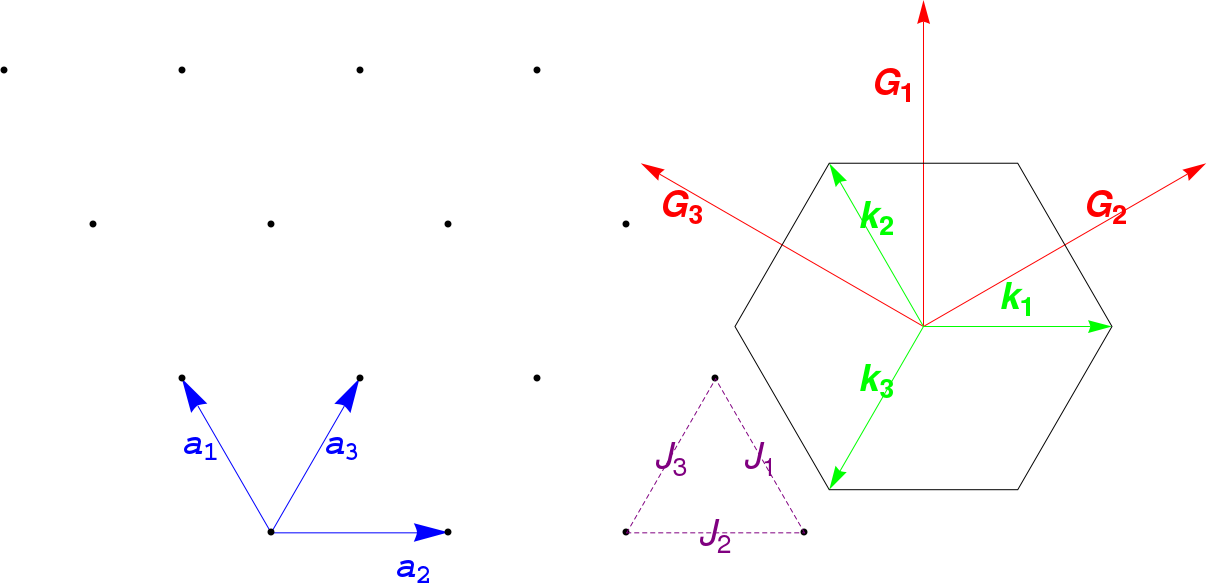}
  \caption{\label{fig:lattice} \CO Triangular lattice in real space (left) and the corresponding reciprocal lattice vectors with the Brillouin zone (right).
  The experimental setup proposed here has a fixed geometry, and the anisotropy of the hopping rates is realized by different intensities of the
  laser beams creating the optical lattice potential.
  Note that in comparison to optical lattices created using standing waves, the lattice spacing is larger, $|\mathbf{a}_i|=2\lambda_L/3$.}
\end{figure}

For concreteness, we consider bosons described by the Bose-Hubbard model
\begin{equation}
 H = -\sum_{<ij>} J_{ij} b^\dagger_i b^{\phantom{\dagger}}_j + \frac{U}{2}\sum_j \hn_j (\hn_j-1) + \sum_j (V_0 \br^2_j -\mu_0) \hn_j \;. \label{eq:def:BHM}
\end{equation}
Here $J_{ij}>0$ describes anisotropic nearest-neighbor hopping between sites $i$ and $j$ on an equiangular triangular lattice (cf. Fig.~\ref{fig:lattice}), 
$U$ is the on-site interaction strength, $V_0$ gives the strength of the external harmonic potential, and the central chemical potential $\mu_0$ sets the total number of particles.

Using numerically robust methods [exact diagonalization (ED), projected entangled pair states (PEPS)~\cite{PEPS} or quantum Monte Carlo (QMC)~\cite{QMC}], it is
very challenging to describe the Bose-Hubbard model in Eq.~(\ref{eq:def:BHM}) with kinetic frustration due to either
the size of the Hilbert space (ED) or the frustration (PEPS, QMC).
We use a low-entanglement (mean-field) approach. We do not expect that it can describe a spin-liquid phase. 
However, the estimates presented here can be used as a starting point to initiate experimental quantum simulations.

This paper is organized as follows. In Sec. II we discuss a specific experimental setup to realize the triangular lattice and outline the band-structure 
calculation. Section III is devoted to the discussion of the experimental parameters based on the setup and the corresponding microscopic
parameters. We also discuss the procedure of reversing the interaction $U$ and harmonic potential $V_0$. In Sec. IV we outline the numerical simulation method.
The results of the simulations are shown and their implications are discussed in Sec. V. 

\section{Triangular optical lattice}

Following Refs.~\cite{shaking-exp,triang-NJP}
we consider three phase stabilized running waves at blue detuning with some wavelength $\lambda_L$ in 120$^\circ$ angles.
The corresponding electric field is given by
\begin{equation}
  \mathbf{E}(\br,t) = \sum_{i=1,2,3} E_i \hat{\epsilon}_i \cos( \bk_i \br - \omega_L t),
\end{equation}
where $E_i$ are the strengths of the electric field in each plane wave, $\omega_L$ is the laser frequency, the wave vectors are
\begin{equation}
 \bk_1 = k_L (1,0) \;,\bk_{2,3} = k_L \left( -\frac{1}{2}, \pm \frac{\sqrt{3}}{2}\right) \;,
\end{equation}
with the wave number $k_L=2\pi/\lambda_L$, and the polarizations $\hat\epsilon_i$ lie in the plane of propagation,
\begin{equation}
 \hat\epsilon_i = k_L^{-1} ( \hat z \times \bk_i) \;.
\end{equation}
The time averaged laser intensity can be reparametrized conveniently as the optical lattice potential
\begin{equation}
 V(\br) = V_{\rm offset} + V_L \sum_{i=1,2,3} (1+\alpha_i) \sin^2( \bb_i \br), \label{eq:latticepot}
\end{equation}
where $\bb_1 = \frac{1}{2} (\bk_2 - \bk_3)$, etc., and $\alpha_i$ characterize the anisotropy of the optical lattice potential. 
For simplicity, we will consider the case of partial anisotropy $\alpha_1 =-\alpha $ and $\alpha_{2,3} = \alpha/2$ corresponding to $E_1 \geq E_2=E_3$~\cite{fullanisotropy}.

The optical lattice setup discussed above has a fixed lattice geometry. This
allows for a direct comparison of the time-of-flight images with different values of $\alpha$.
Rhombic optical lattices can also
be realized with two standing waves by varying the angle between the beams. 
However, it is harder to reach the isotropic case (which follows trivially from symmetry with the three-beam setup)
and the comparison of TOF images is not so straightforward as the reciprocal lattice vectors change.
An additional experimental advantage of the setup proposed here over a two-beam setup is that the 
laser intensities can be changed more easily than the angle between the beams.

The periodic potential in Eq.~(\ref{eq:latticepot}) defines (possibly overlapping) Bloch bands for a triangular lattice.
The lattice vectors $\mathbf{a}_{1} = (-1/3,1/\sqrt{3}) \lambda_L$, $\mathbf{a}_{2} = (2/3,0) \lambda_L$, and $\mathbf{a}_{3} =\mathbf{a}_{1} + \mathbf{a}_{2}$ are shown in Fig.~\ref{fig:lattice}.
The band-structure parameters for Eq.~(\ref{eq:def:BHM}) are calculated using the solution of the 
two-dimensional one-body problem in the optical lattice potential~(\ref{eq:latticepot}):
for each fixed momentum $\bk$ of the Brillouin zone (BZ), we calculate the eigenvalues and
eigenvectors of the block matrix (note rescaling in terms of the recoil energy $E_R=\hbar^2k_L^2/2M$)
\begin{equation}
  h_{\tilde \bk,\tilde \bk'} = \delta_{\tilde \bk',\tilde \bk} \;  \frac{{\tilde k}^2}{k_L^2}  - \frac{V_L}{4E_R} \sum_{j=1}^3 (1 + \alpha_j)[\delta_{\tilde \bk',\tilde \bk+{\mathbf G}_j} + \delta_{\tilde \bk',\tilde \bk-{\mathbf G}_j} ] \label{eq:band-eigen}
\end{equation}
where the extended momentum $\tilde \bk = \bk + g_1 {\mathbf G}_1+ g_2 {\mathbf G}_2$ can be indexed by the integers $g_{1,2} \in [-g_c, g_c-1]$ and $ {\mathbf G}_j =2\mathbf{b}_j$. 
The lowest eigenvalues of Eq.~(\ref{eq:band-eigen}) as a function of $\bk$ define the lowest Bloch band, which is approximately a nearest-neighbor dispersion relation,
\begin{equation}
  \epsilon_{\bk} = -2 \sum_{j=1,2,3} J_j \cos( \bk \cdot \mathbf{a}_j ), \label{eq:def:epsilonk}
\end{equation} 
for the parameter values of $V_L$ and $\alpha$ relevant to this work. Due to the partial potential anisotropy, there is partial hopping anisotropy $J_1=J_3 > J_2$ for $\alpha > 0$.

The minimum of the noninteracting dispersion relation $\epsilon_\bk$ is always
at $\bk=0$ momentum, while the \emph{maxima} lie at
\begin{equation}
\bk_{A,B} =\left( \pm k^*, \frac{\sqrt{3}}{2} k_L \right); \frac{k^*}{k_L} = \frac{3}{2\pi} {\rm arccos}\left(\frac{J_1}{2J_2}\right) . \label{eq:kstar} 
\end{equation}
These points coincide with the corners of the Brillouin zone in the isotropic case, $J_1=J_2=J_3$ ($\alpha=0$). The momenta are not equivalent in terms of
modulo reciprocal lattice vectors, $\bk_A \ncong \bk_B \cong - \bk_A$, leading to the possibility of the $Z_2$ (chirality) symmetry breaking~\cite{shaking-exp,shaking-exp2}. The vectors $\bk_{A,B}$ are incommensurate with the lattice for a general $\alpha$.
The value of $k^*$ decreases for $\alpha > 0$ and vanishes at the rhombic transition point $\alpha=\alpha_c$ when $J_1 = 2 J_2$ , thus $k^*$ serves also as a measure for frustration.
For stronger anisotropy, the lattice links with the stronger hopping define a rhombic lattice. Similar to the square lattice, the bare rhombic lattice is bipartite and therefore there is no frustration.

The joint set of elements of all eigenvectors  $v_{\tilde \bk}$ of Eq.~(\ref{eq:band-eigen}) of
the lowest band define the \emph{Fourier components} of the Wannier function up to a phase factor, 
\begin{equation}
 w(\tilde \bk) \sim v_{\tilde \bk} \to w_j(\br) = \sum_{\tilde \bk} e^{i \tilde \bk (\br - \br_j)} w(\tilde \bk), \label{eq:def:wk}
\end{equation}
which is used on one hand to calculate the envelope for the time-of-flight (TOF) images and to define the dimensionless interaction overlap integral 
\begin{equation}
 u = k_L^{-2} \int\!\!d^2\br \; | w_0(\br)|^4 \;. \label{eq:def:u}
\end{equation}

\section{Experimental and model parameters}

We consider blue detuned laser beams at wavelength $\lambda_L = 736.65$ nm for $^{39}$K atoms, which was 
also used in the experiment described in Ref.~\cite{negT-exp}. 
This implies that the recoil energy is 
\begin{equation}
 E_R = \frac{\hbar^2 k_L^2}{2M} \approx k_B \, 450 \, \mathrm{nK} \approx 2\pi\hbar\, 9.4 \, \mathrm{kHz}.
\end{equation}
Hopping amplitudes and other band parameters in the isotropic case are shown in Table~\ref{tbl:params-isotropic} and for the anisotropic case in Tables~\ref{tbl:params-s2} and  \ref{tbl:params-s3}.

\begin{table}
 \caption{\label{tbl:params-isotropic} Band parameters in the isotropic case ($\alpha=0$) for different values of the lattice depth $V_L$. The bandwidth of the lowest band $\epsilon_\bk$ is $W$, and $\Delta$ gives the gap between the lowest Bloch band and the next one (or the bottom of the continuum).
 The next-nearest-neighbor hopping is $J_{\rm nnn}$.
 Note that the hopping rates vanish faster with increasing $V_L$ than in a standing wave optical lattice; cf. Table I in Ref.~\cite{Blochreview}. The main reason
 is the larger lattice spacing.
 }
 \centering
\begin{tabular}{llllll}
 $V_L/E_R$ & $W/E_R$ & $\Delta/E_R$ & $J_1/E_R$ & $\Delta/W$ & $J_{\rm nnn}/J_1$ \\
\hline
 1. &  0.529 & 0.572 & 0.0627024 &  1.08 &  -0.08297\\
 2. &  0.257 & 1.412 & 0.0295701 &  5.59 & -0.03648\\
 3. &  0.116 & 2.233 & 0.0133173 &  19.28 & -0.01476\\
 4. &  0.0548 & 2.975 & 0.00625151 &  54.26 & -0.00627\\
 5.5 &  0.0195 & 3.903 & 0.00221681 &  199.9 & -0.00197\\
\end{tabular}
\end{table}

\begin{table*}
 \caption{\label{tbl:params-s2} Band parameters for different values of the lattice potential anisotropy $\alpha$ for $V_L=2E_R$. }
 \centering
\begin{tabular}{lllllllll}
 $\alpha$ & $W/E_R$ & $\Delta/E_R$ & $J_1/E_R=J_3/E_R$ & $J_2/E_R$ & $\Delta/W$ & $J_1/J_2$ & $k^*/k_L$ & $u (\sim U)$ \\ 
 \hline
  0. & 0.257 & 1.412 & 0.0295701 & 0.0295701 & 5.586 & 1. & 0.5 & 0.167994 \\
  0.25 & 0.262 & 1.277 & 0.0320253 & 0.0254546 & 4.871 & 1.25813 & 0.425 & 0.167238\\
  0.5 & 0.275 & 1.145 & 0.0347427 & 0.022295 & 4.166 & 1.55832 & 0.323 & 0.165063\\
  0.75 & 0.296 & 1.01 & 0.0377333 & 0.0199259 & 3.408 & 1.89368 & 0.156 & 0.161633\\
  1. & 0.325 & 0.873 & 0.0409952 & 0.0182272 & 2.689 & 2.24912 & 0. & 0.157101\\
\end{tabular}
\end{table*}

\begin{table*}
 \caption{\label{tbl:params-s3} Band parameters for different values of the lattice potential anisotropy $\alpha$ for $V_L=3E_R$. 
  The first column defines the identifiers for the different ``protocols''.
 }
 \centering
\begin{tabular}{llllllllll}
 & $\alpha$ & $W/E_R$ & $\Delta/E_R$ & $J_1/E_R=J_3/E_R$ & $J_2/E_R$ & $\Delta/W$ & $J_1/J_2$ & $k^*/k_L$ & $u(\sim U)$\\ 
 \hline
 a) & 0. & 0.116 & 2.233 & 0.0133173 & 0.0133173 & 19.28 & 1. & 0.5 & 0.226388\\
 b) & 0.1 & 0.118 & 2.156 & 0.0139504 & 0.0121748 & 18.28 & 1.14585 & 0.459 & 0.226222\\
 c) & 0.2 & 0.121 & 2.079 & 0.014631 & 0.0111764 & 17.24 & 1.3091 & 0.409 & 0.225722\\
 d) & 0.3 & 0.124 & 2.003 & 0.015362 & 0.0103051 & 16.16 & 1.49072 & 0.348 & 0.224902\\
 e) & 0.4 & 0.128 & 1.926 & 0.0161468 & 0.0095464 & 15.08 & 1.6914 & 0.269 & 0.223772\\
 f) & 0.5 & 0.134 & 1.847 & 0.0169886 & 0.008888 & 13.74 & 1.91141 & 0.143 & 0.222343\\
 g) & 0.6 & 0.142 & 1.768 & 0.0178908 & 0.00831952 & 12.48 & 2.15046 & 0. &  0.220625\\
 h) & 0.7 & 0.150 & 1.689 & 0.0188569 & 0.00783222 & 11.30 & 2.40761 & 0. & 0.218627 \\
\end{tabular}
\end{table*}

The external harmonic potential has a bare strength $\bar V$,
\begin{equation}
 V_0/E_R \equiv \pm \bar V \nu^2  \approx \pm 2.78 \times 10^{-8} \; \nu^2 ,
\end{equation}
where the value of the trapping frequency $\nu$ is in units of Hz. The upper sign corresponds to the usual trapping potentials, the lower sign
is active for the antitrapping situation. 

The on-site interaction is given by~\cite{Blochreview}
\begin{equation}
 U/E_R = 8\pi \; (a_s k_L) \; u \; w_z ,
\end{equation}
where, for simplicity, we input the value of the scattering length $a_s$ directly~\cite{asB}. The interaction overlap $u=u(V_L,\alpha)$  is calculated from the Wannier function in Eq.~(\ref{eq:def:u}). 
We consider a layered system  similar to Ref.~\cite{negT-tdGA} with a vertical optical lattice depth $V_{L,\rm ver} = 25 E_R$~\cite{verticalbeam}, which corresponds to a vertical hopping $J_z \approx 0.00104 E_R$ and Wannier overlap $w_z\approx0.848035$.

To reach negative absolute temperatures in Ref.~\cite{negT-exp}, the experimental parameters (horizontal and vertical
optical lattice intensities, magnetic field, etc.) were changed via an involved protocol. 
However, most of these steps emerge as a technical necessity. 
Furthermore, the cloud is initially trapped in a very deep optical
lattice where the atomic density distribution is essentially frozen. 
From this perspective, most steps of the experimental protocol are almost instantaneous.

To simplify the numerical simulations and to improve the transparency of the text, we consider an \emph{instantaneous quench} in the system:
for time $t < 0$, we take an isotropic ($\alpha_i=0$) triangular lattice with depth $V_{L,i}=5.5 E_R$, a scattering length $a_{s,i}= +400 a_{\rm Bohr} (U/J_1 \approx 582)$ and $\nu_i = 60$ Hz horizontal trapping frequency ($V_0/E_R \approx 0.0001 $) for a strongly compressed Mott insulator initial ground state in equilibrium. 
According to Ref.~\cite{negT}, such an initial state is necessary to optimize the final condensate fraction.
At $t=0$, we instantaneously change to a shallower optical lattice $V_{L,f} < V_{L,i}$, a negative scattering length $a_{s,f} < 0$ and an anti-trapping harmonic potential $V_{0,f}<0$~\cite{negativeV0U}.

The optimal regime of the final lattice depth for the numerical and the experimental setups depends on various aspects. 

Fast enough dynamics certainly requires weak enough $V_{L,f}$. Avoiding technical heating from the blue detuned lattice lasers also favors weaker lattice potentials. 

On the other hand, there are more arguments in favor of a relatively deep lattice. If the lattice is too weak, the Bloch gap between the lowest band and the next band may not be large enough (cf. Tables~\ref{tbl:params-s2} and \ref{tbl:params-s3}). This is unfavorable since the protocol strongly relies on the bounded kinetic energy, which is violated if the Landau-Zener tunneling rate to other Bloch bands is not negligible.
In weaker lattices the overlap integral for the interaction is also reduced and therefore larger scattering lengths are needed to compensate. This usually implies getting closer to a Feshbach resonance~\cite{negT-exp}, where many-body losses are enhanced. 
Last, the value of the lattice potential anisotropy is bounded, $\alpha \leq 1$, since the wave intensities cannot be negative; cf. Eq.~(\ref{eq:latticepot}).
Additionally, for $V_L=3E_R$, the rhombic transition happens at a lower value of $\alpha$ than for $V_L=2E_R$ (cf. Tables~\ref{tbl:params-s2} and \ref{tbl:params-s3}), which might be favored experimentally.

Taking these considerations into account, we will mainly focus on the parameters $V_{L,f}=3 E_R$, a scattering length $a_{s,f}= -50 \, a_{\rm Bohr}$, and anti-trapping $\nu_f = 30$ Hz ($V_0/E_R \approx - 0.000025 $). 
For these parameters the system is well approximated by the one-band Hubbard Hamiltonian in Eq.~(\ref{eq:def:BHM}).

\section{Time-dependent Gutzwiller Ansatz}

We apply the time-dependent Gutzwiller ansatz (GA) \cite{negT-tdGA,TDG-1,TDG-2,TDG-3,TDG-4,TDG-5,TDG-6} to study the dynamics of the cloud after the quench defined in the previous section. In this approximation, the probability amplitudes of finding precisely $m$ bosons at site $j$ and time $t$ are given by the following set of differential equations:
\begin{eqnarray}
 i \partial_t f_m(j,t) &=& [ U(t) \frac{m-1}{2} + V_0(t) \br_j^2 - \mu_0 ] m\, f_m(j,t) \nonumber \\
    &&  - \Phi^*(j,t) \, \sqrt{m+1} \, f_{m+1}(j,t) \nonumber \\
    && - \Phi(j,t) \, \sqrt{m}  \, f_{m-1}(j,t) , \label{eq:def:tdG}
\end{eqnarray}
where we introduced $\Phi(j,t) = \sum_{\delta} J_{\delta}(t) \langle b_{j+\delta} \rangle$, the index $\delta$ running over
the six nearest-neighbor sites, and
\begin{eqnarray}
\langle b_j \rangle &=& \sum_m \sqrt{m+1} f_m^*(j,t) f_{m+1}(j,t) . 
\end{eqnarray}

In the GA, quantum correlations beyond the mean-field $\Phi$ between the lattice sites are neglected. In higher dimensions, or more precisely, for higher coordination numbers $z$, the approximation is expected to improve. For example, for the cubic lattice with $z=6$, the GA variational wave function gives a good estimate for the quantum phase transition between the Mott insulator and the superfluid phase~\cite{Blochreview}.
In Ref.~\cite{negT-tdGA} we studied numerically a setup corresponding to the experiments in Ref.~\cite{negT-exp} on the square lattice, $z=4$. Based on these findings, deep in the rhombic regime $J_1 \gg J_2$, the time-dependent GA should work reasonably well.
The isotropic triangular lattice with $z=6$ is closer to the mean-field limit $z\to\infty$ than the square or rhombic lattice; however, frustration is expected to enhance quantum fluctuations which are captured poorly in the mean-field approximation. 
Nevertheless, we will confirm later that the dynamics in the GA gives the expected behavior in the isotropic limit.
Similar to  Ref.~\cite{negT-tdGA}, we focus only on a single layer and entirely neglect the hopping between layers.

The lattice consists of $192\times 192$ lattice sites. The initial state is a strongly compressed Mott insulator, which is determined as the ground state of Eq.~(\ref{eq:def:BHM}) in the equilibrium variational GA for the initial parameters. The total atom number is $N_{\rm tot} \approx 2260$. 
To numerically integrate Eq.~(\ref{eq:def:tdG}) we use the fourth-order Runge-Kutta method and the (conserved) total atom number
\begin{eqnarray}
N_{\rm tot} &=& \sum_{j}  n_j(t), \textrm{with} \label{eq:def:Ntot} \\
 n_j(t)  &=& \sum_m \, m \, |f_m(j,t)|^2\nonumber, 
\end{eqnarray}
serves as a primary measure of numerical accuracy.

We note that after the quench the total energy $E_{\rm tot} = \langle\lbrace f_m(j,t) \rbrace \vert H_f \vert \lbrace f_m(j,t) \rbrace \rangle + \mu_0 N_{\rm tot}$
is also conserved. The $N_{\rm tot}$ and $E_{\rm tot}$ determine a unique grand-canonical density matrix for the Hamiltonian Eq.~(\ref{eq:def:BHM}),
and in principle the long-time averages of (macroscopic) quantities in the GA should approximate the corresponding expectation values.
However, at the moment it is unclear how the latter could be computed.

\section{Numerical results }

We calculate various (macroscopic) quantities as a function of time, as defined in Ref.~\cite{negT-tdGA}. 
The total pair density is given by
\begin{eqnarray}
D_{\rm tot} &=& \sum_{j}  d_j(t), \textrm{with} \nonumber \label{eq:def:Dtot} \\
 d_j(t)  &=& \sum_m \, \binom{m}{2} \, |f_m(j,t)|^2\nonumber .
\end{eqnarray}
The time evolution of the condensate occupation
\begin{eqnarray}
N_0(t) &=& \sum_{j}  | \langle b_j \rangle |^2, \label{eq:def:N0}
\end{eqnarray}
and nearest-neighbor coherences 
\begin{equation}
C(t) =  \sum_{j,\delta} \langle b_{j}^\dagger b_{j+\delta}^{\phantom{\dagger}} \rangle \overset{GA}{=}  \sum_{j,\delta} \langle b_{j}^\dagger \rangle \langle b_{j+\delta}^{\phantom{\dagger}} \rangle \label{eq:def:C}
\end{equation}
follow a qualitatively similar behavior as on the square lattice~\cite{negT-tdGA}, $C$ becoming
negative; see Fig.~\ref{fig:iso-macros}. 
Here and below $GA$ means that the expectation values are evaluated using the bosonic Gutzwiller wave function.

Longer-range coherences ($l \neq 0$) are calculated in the $\mathbf{a}_2$ ($x$) direction,
\begin{eqnarray}
  {\texttt C}(l,t) &=& \sum_j \langle b_{\br_j}^\dagger b_{\br_j+ l \mathbf{a}_2} \rangle \overset{GA}{=} \sum_j \langle b_{\br_j}^\dagger  \rangle  \langle b_{\br_j+ l \mathbf{a}_2} \rangle , \label{eq:def:Cl} \\
  \tilde{\texttt C}(l,t) &=&\sum_j \frac{ \langle b_{\br_j}^\dagger b_{\br_j+ l \mathbf{a}_2} \rangle }{\sqrt{ n(\br_j) n(\br_j+ l \mathbf{a}_2)}} \nonumber \\
  &\overset{GA}{=}& \sum_j \frac{ \langle b_{\br_j}^\dagger  \rangle }{\sqrt{ n(\br_j)}} \frac{ \langle b_{\br_j+ l \mathbf{a}_2} \rangle}{\sqrt{ n(\br_j+ l \mathbf{a}_2)}}, \label{eq:def:Ctildel}
\end{eqnarray}
the latter being normalized so that it is less sensitive to spatial inhomogeneities.

We calculate two-dimensional TOF images  using the formula (following Ref.~\cite{Blochreview})
\begin{equation}
 I_{\rm TOF}(\tilde \bk,t) = |w(\tilde \bk)|^2 \; {\cal G}(\tilde \bk,t) \label{eq:def:TOF}
\end{equation}
where the envelope $|w(\tilde \bk)|^2$ is the Fourier transform of the Wannier function (c.f. Eq.~(\ref{eq:def:wk})) and the Fourier transform of the single-particle density matrix at a time $t$ is given in the GA by
\begin{eqnarray}
 {\cal G}(\bk,t) & \overset{GA}{=} & |\langle b_{\bk}(t) \rangle |^2 + L^{-2} (N_{\rm tot} - N_0(t)) \nonumber \\
 \langle b_{\bk} \rangle &=& L^{-1} \sum_j e^{i\bk \br_j} \langle b_j \rangle , L=192.
\end{eqnarray}
This normalization implies $\sum_{\bk \in BZ}  {\cal G}(\bk, t) = N_{\rm tot} $, and makes
direct comparisons of the absolute TOF intensities possible. 
While the different expectation values in Eqns.~(\ref{eq:def:N0})-(\ref{eq:def:Ctildel}) provide valuable insight 
regarding certain quantities and the structure of correlations at various times on the mean-field level, 
the TOF intensities calculated using Eq.~(\ref{eq:def:TOF}) can be compared directly to experiments.

\subsection{Numerical results for the isotropic lattice}

\begin{figure}
\centering
 \includegraphics[width=0.48\textwidth,clip=true]{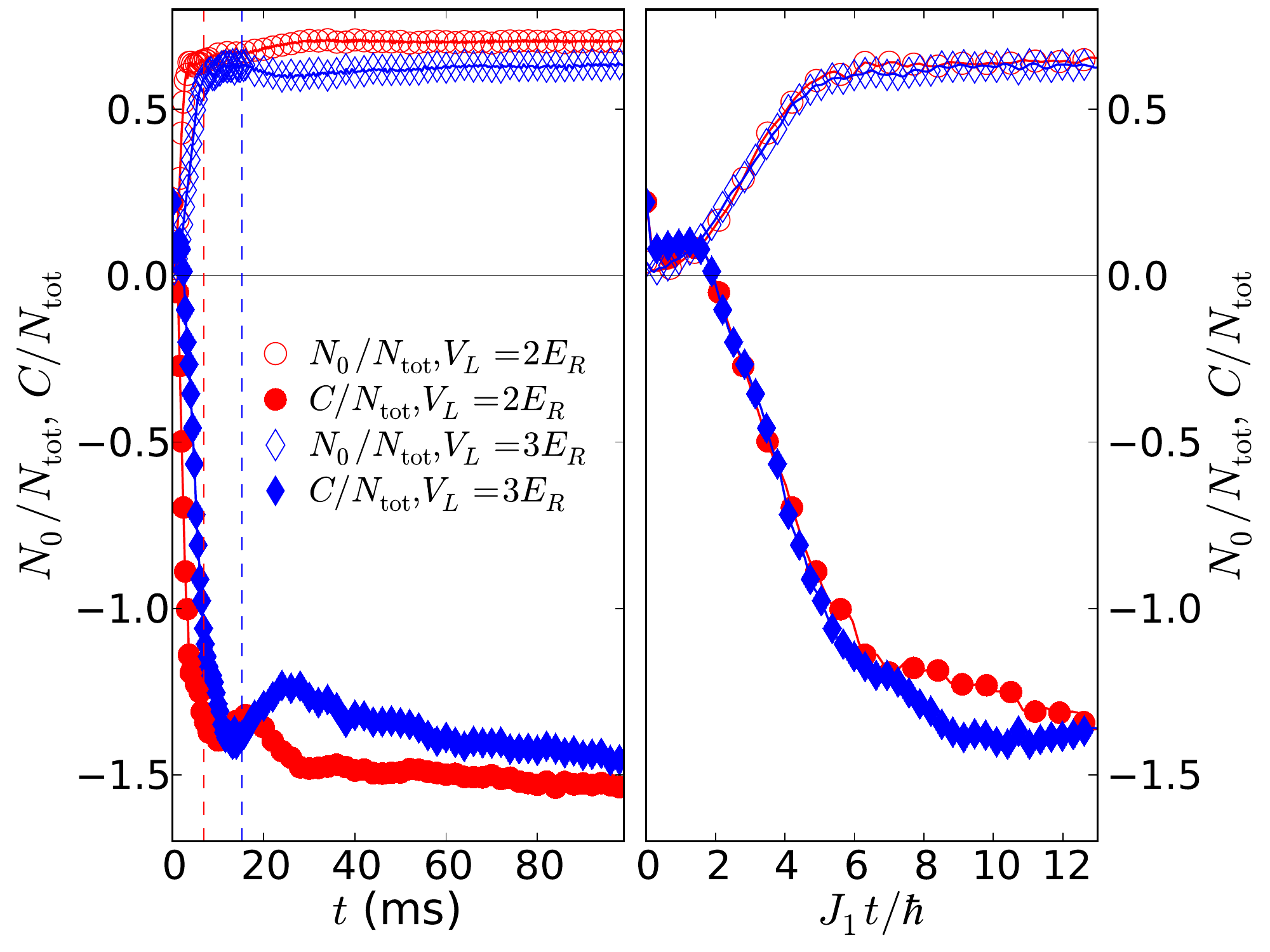} \\
 \caption{\label{fig:iso-macros} \CO
  Left: condensate occupation
  $N_0(t)$ and nearest-neighbor coherences $C(t)$ as a function of time $t$ in the isotropic case $\alpha=0$ for two different final lattice depths. 
  Right: the same quantities with the time axis rescaled by $J_1 /\hbar$. While for $V_L=3E_R$ the hopping amplitude
  is lower (cf. Table~\ref{tbl:params-isotropic}), and therefore the dynamics is slower in real time, the initial 
  evolution in the natural time unit $\sim J_1 t$ is the same as for $V_L=2E_R$. 
  The dashed lines on the left panel correspond to $J_1 t/\hbar=12$.
 }
\end{figure}

\begin{figure*}
\centering
 \includegraphics[width=0.95\textwidth,clip=true]{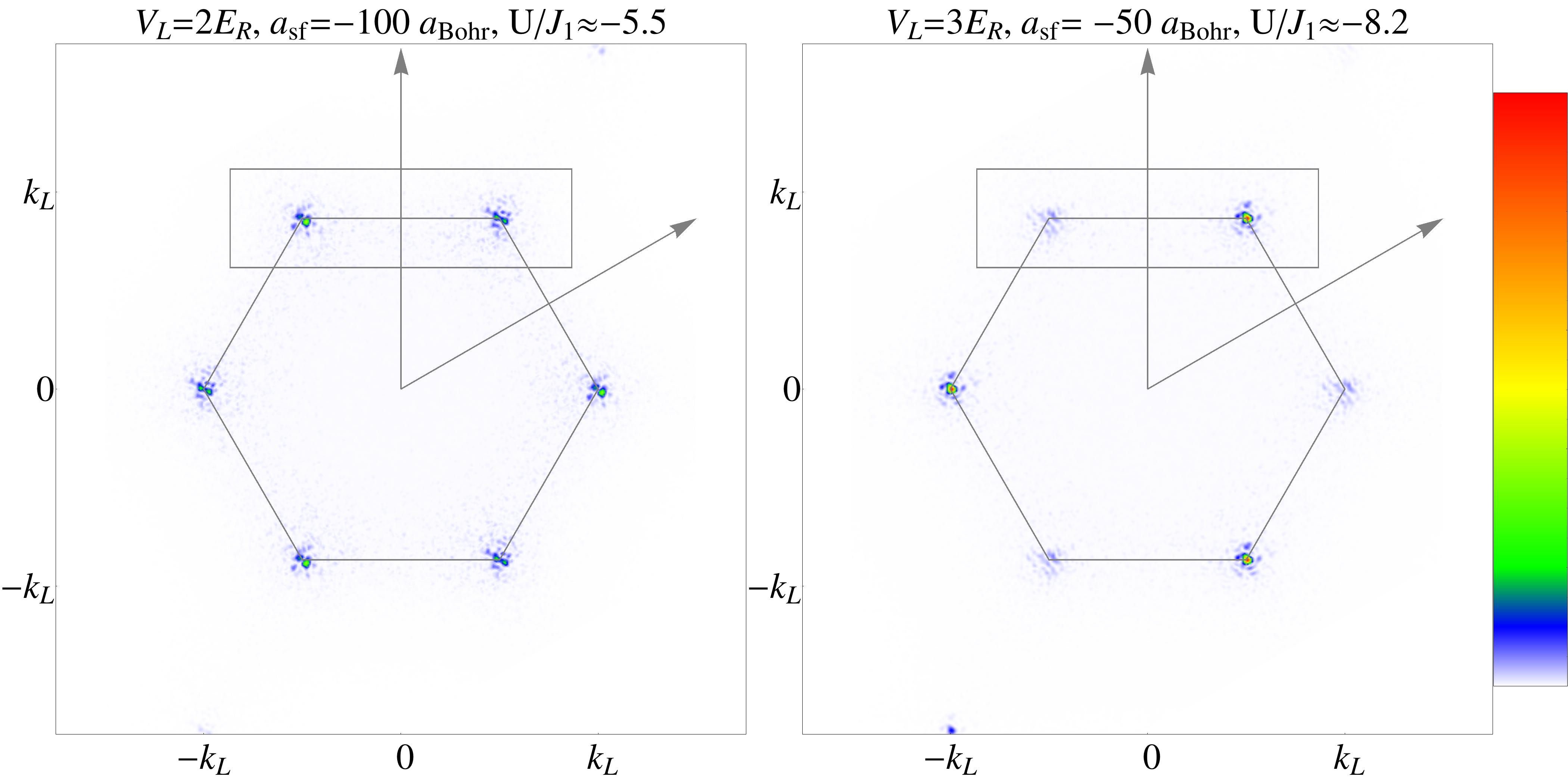}
 \caption{
 \label{fig:iso-tof} \CO
  TOF images at $t=200$ ms for a weaker ($V_L/E_R=2$, $U/J_1 \approx - 5.5$) and a stronger final lattice ($V_L/E_R=3$, $U/J_1 \approx - 8.2$) in the isotropic case. 
  Note that the latter image shows a strong chirality.
  The arrows represent the reciprocal lattice vectors $\mathbf{G}_1$ and $\mathbf{G}_2$, the hexagon represents the border of the first BZ. The rectangular area is displayed in Fig.~\ref{fig:bqpart}.
 }
\end{figure*}

We compare macroscopic quantities for two different sets of the final parameters, $V_{L,f}=2E_R, a_{s,f}=-100 a_{\rm Bohr}$ and  $V_{L,f}=3E_R, a_{s,f}=-50 a_{\rm Bohr}$ in Fig. \ref{fig:iso-macros}.
The interaction strengths are $U/J_1 \approx - 5.5$ and $U/J_1 \approx - 8.2$, respectively. 
The total energies after the quench are $E_{\rm tot}(V_{L,f}=2E_R)/E_R \approx -46.31$ and $E_{\rm tot}(V_{L,f}=3E_R)/E_R \approx -38.15$.
TOF images at $t=200$ ms are shown for comparisons in Fig.~\ref{fig:iso-tof}.
The main contribution of the TOF intensities is concentrated around the corners of
the Brillouin zone. These peaks persist over time in contrast to the noisy features
representing spatial variations, which also change as a function of time and thus
would cancel out over averaging.
The noisy features in the TOF images are likely the results of the fact
that the system is in a non-equilibrium state after the quench. How
exactly these noisy features develop is an open question.


Since in the case of the deeper lattice $V_{L,f}=3E_R$ the system is closer to the hard-core limit and yet the corresponding TOF images
show more enhanced peaks with a pronounced chirality, we will consider this lattice depth in the following.

\subsection{Numerical results for the anisotropic lattice}

\begin{figure}
  \centering
 \includegraphics[width=0.45\textwidth]{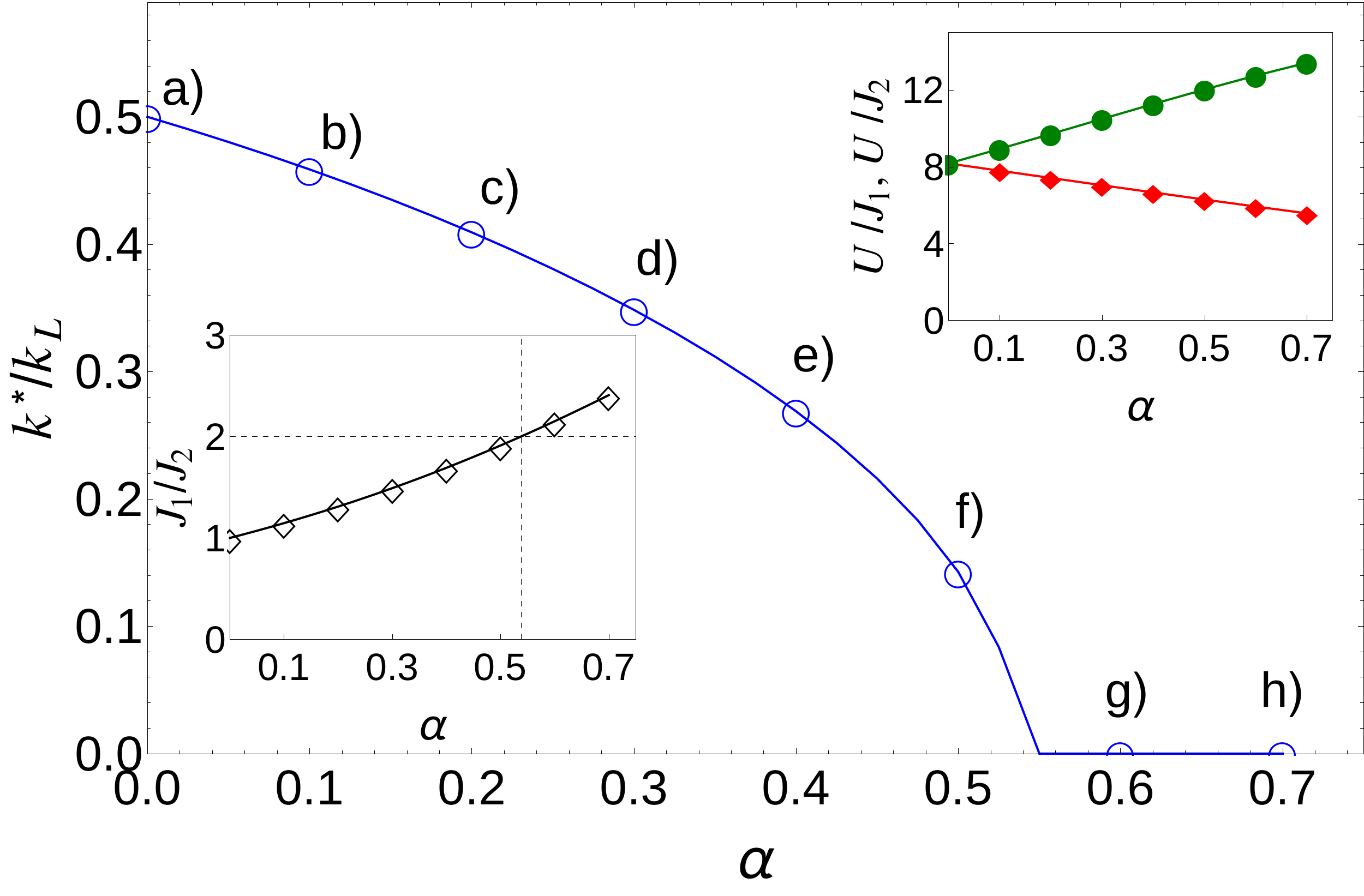}
 \caption{\label{fig:micros} \CO Value of $k^*$ as defined by Eq.~(\ref{eq:kstar}) as a function of the optical potential anisotropy $\alpha$ for $V_L=3E_R$.
 Insets: microscopic parameters for the different values of $\alpha$ listed in Table~\ref{tbl:params-s3}. The rhombic transition point corresponds to
 $\alpha = \alpha_c \approx 0.55$. }
\end{figure}

In this subsection we display macroscopic quantities and TOF images for different values of the potential anisotropy $\alpha$ for the final lattice depth $V_{L,f}=3E_R$. The microscopic parameters used for the simulations
are listed in Table~\ref{tbl:params-s3} and shown in Fig.~\ref{fig:micros}.

\begin{figure}
  \centering
 \includegraphics[width=0.45\textwidth]{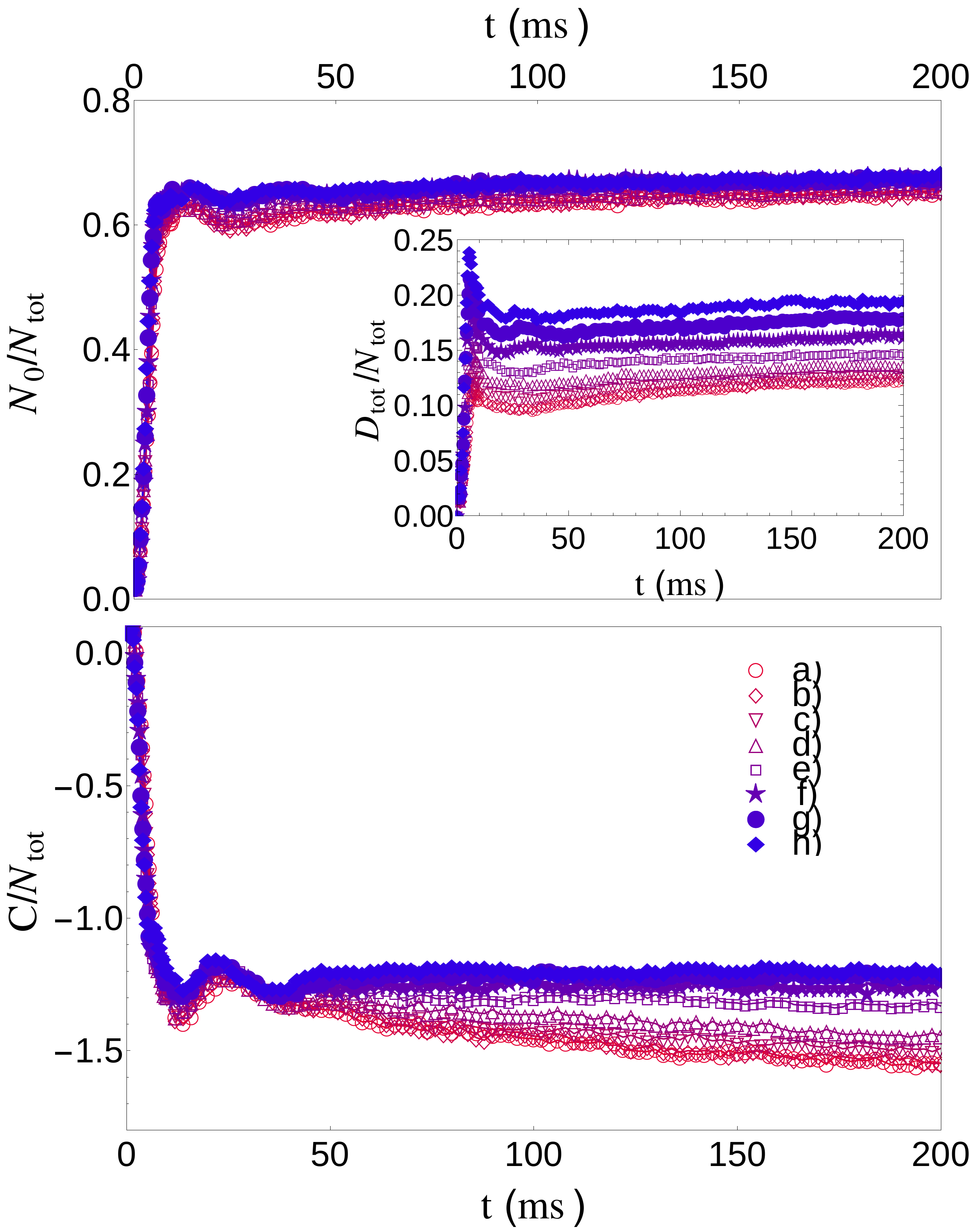}
 \caption{\label{fig:macros} \CO Macroscopic parameters 
 (condensate occupation $N_0$, total pair density $D_{\rm tot}$, and nearest-neighbor coherences $C$) 
 as a function of time $t$ for the different values of $\alpha$ listed in Table~\ref{tbl:params-s3}. Note that
 the condensate occupation $N_0$ varies very weakly with $\alpha$.}
\end{figure}

The condensate occupations $N_0$, the nearest-neighbor coherences $C$ and the total pair density $D_{\rm tot}$ is shown in Fig.~\ref{fig:macros} as a function of time $t$ for the different values of $\alpha$.

\begin{figure*}
 \centering
 \includegraphics[width=0.95\textwidth]{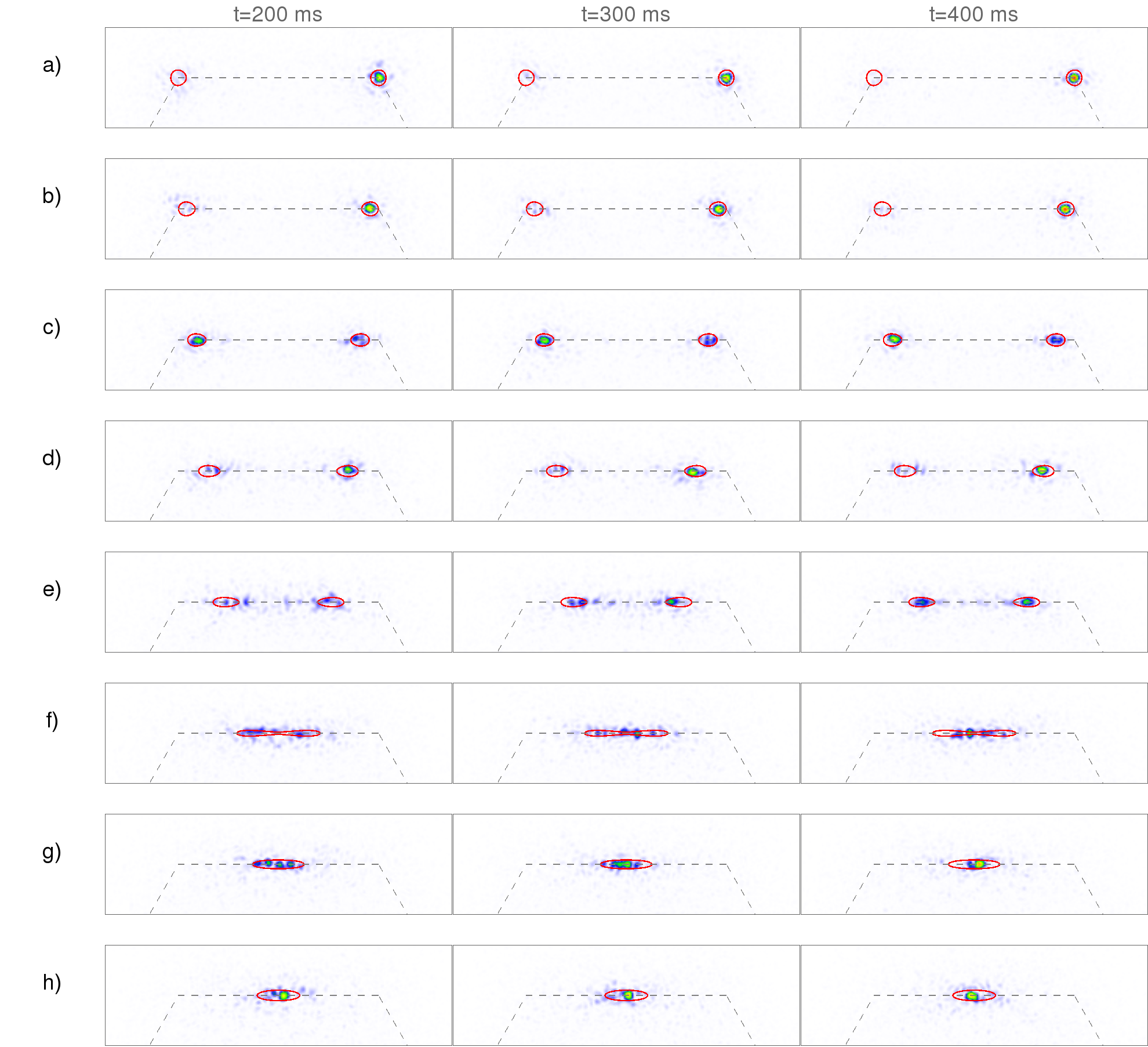}
 \caption{\label{fig:bqpart} \CO Parts of the TOF images at  $V_{L,f}=3E_R$ for different values of $\alpha$. The rows (a) - (h) correspond to 
 different values of the lattice potential anisotropy $\alpha$ (cf. Table~\ref{tbl:params-s3} and Fig.~\ref{fig:micros}); the columns correspond to different times after the quench. The color coding is similar to Fig.~\ref{fig:iso-tof}.
 The gray dashed line corresponds to the border of the BZ. 
 The red contours are defined to enclose 128 points with the highest kinetic energy $\epsilon_\bk$ in reciprocal space
 for the lattice size of $192\times192$.  
 }
\end{figure*}

We show the relevant part of the TOF images for the different values of $\alpha$ in Fig.~\ref{fig:bqpart}.
For weak anisotropies $\alpha \leq 0.4$, we observe coherence manifesting in peaks at the corresponding 
quasiclassical maxima $\bk_A$ or $\bk_B$ of the free dispersion. 
The asymmetry in the TOF intensities at the two momenta $\bk_A$ and $\bk_B$ is related to the chirality when $k^* \neq 0$.
Beyond the rhombic transition point, $2 J_2 < J_1$ corresponding to
$\alpha > 0.55$, there is also apparently  coherence at the ``N\'eel'' momentum $\mathbf{Q}=(0,\sqrt{3}/2)k_L$. However, for the intermediate value $\alpha = 0.5$ [ case (f)], there are no peaks at the quasiclassical maxima; furthermore, no dominant coherent peak is found up to $t=600$ ms of the numerical simulation.
Note that the macroscopic quantities $N_0, C$, or $D_{\rm tot}$ show a weak monotonous behavior as a function of $\alpha$, cf. Fig.~\ref{fig:macros}.

\begin{figure*}
 \centering
 \includegraphics[width=0.95\textwidth]{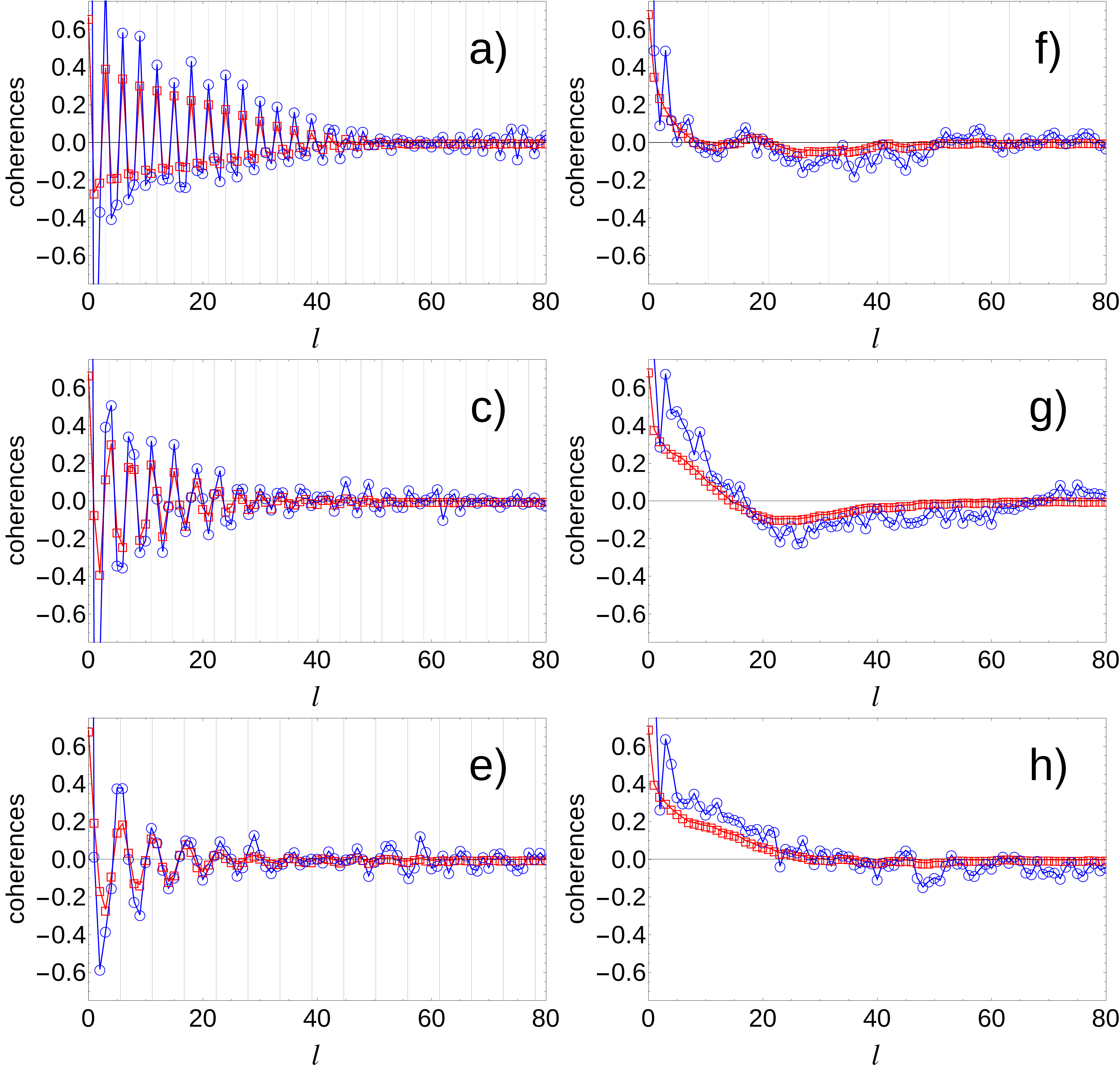}
 \caption{\label{fig:CCtilde} \CO Longer-range coherences in the $\mathbf{a}_2$ (x-) direction as a function of the site index $l$ at time $t=400$ ms. Red squares: ${\rm Re} \, {\texttt C}(l) $; blue circles: ${\rm Re} \, \tilde {\texttt C}(l) $. The grid lines represent the quasiclassical ``pitch'' length $2\pi/|\mathbf{a}_2|k^*$ of the spiral ordering vector.
  }
\end{figure*}

To complement the TOF images, we calculated longer-ranged coherences, shown in Fig.~\ref{fig:CCtilde}.
While the density normalized coherences $\tilde{\texttt C}$ are more noisy, they follow qualitatively the course of the corresponding unnormalized coherences, which in turn are well approximated in most cases by 
\begin{equation}
 {\texttt C}(x) \approx c(x) e^{i \kappa x} \;, \kappa = k^* + i \xi^{-1} \;, \label{eq:coherencelengths}
\end{equation}
and $c(x)$ varies weakly.
As a  sidenote, $\xi$ determines the approximate size of the coherent ``domains,'' each contributing with a given chirality $\pm$. The asymmerty between 
$\mathbf{k}_A$ and $\mathbf{k}_B$ becomes suppressed as $\xi$ decreases. This can be observed in the case e) or at $V_L/E_R=2$ in the isotropic case.
The notable exception to the behavior given by Eq.~(\ref{eq:coherencelengths})
is the case (f), without any apparent oscillating component and a coherence length $\xi$ of the order of one lattice spacing.
From Eq.~(\ref{eq:coherencelengths}) it is obvious that coherence cannot be defined if the condition for coherent behavior
\begin{equation}
 k^* \xi \gg 1
\end{equation}
breaks down, i.e., when the ``pitch length'' of the spiral in the x-direction, $\sim 1/k^*$, becomes longer than the coherence length $\xi$.
Since $k^* = k^*(\alpha) \sim \sqrt{\alpha_c-\alpha} \to 0$ at the rhombic transition, it is important how $\xi$ depends on the various parameters and
cloud size around $\alpha \approx \alpha_c$. 
Unfortunately, this question cannot be addressed properly using the present approach. 

An obvious reason for the lack of coherence in the case (f)
could be an anomalously long ``relaxation'' time due to the various approximations, 
and the experimental system [or even the ``true'' dynamics under Eq.~(\ref{eq:def:BHM})] could display coherence in shorter times. 
However, this claim is only partially justified. It is true that within the time-dependent GA defined by Eq.~(\ref{eq:def:tdG}) the ``relaxation'' is slow as it is driven by \emph{dephasing}, i.e., by the mismatch between local mean-field Hamiltonians. 
The true dynamics governed by Eq.~(\ref{eq:def:BHM}) should lead to a faster \emph{equilibration} in general. However, the main candidate state
for stronger on-site interactions has simply N\'eel-type coherence~\cite{shaking-EPL,XY-MSWT}, which is found to develop for $J_1 > 2J_2$. 
Therefore, it is quite puzzling why GA fails to find either of the mean-field type solutions (spiral or N\'eel) in the case (f). 

Regarding time scales, one should also not forget about \emph{losses and decoherence} in the experimental system, which includes all processes that are left out from Eq.~(\ref{eq:def:BHM}): technical heating from the lasers, multiband contributions, many-body losses driven by three-body recombination, etc. These determine the experimental lifetime and provide an upper bound to the coherence lifetime. For the optimal square-lattice setup it was found to be on the order of 700 ms~\cite{negT-exp}.

\section{Conclusions}

We proposed a specific experimental setup for interacting bosons on an anisotropic triangular lattice and calculated microscopic parameters for the corresponding Bose-Hubbard model.                                                                                
We studied  numerically the dynamics of the atoms in a time-dependent mean-field approximation after instantaneously reversing the signs of the on-site interaction to $U_{f} <0$ and the external potential to $V_{0,f} <0$. We found that quasi-classical coherence with 120$^\circ$ spiral order develops in the isotropic case. This can be interpreted as a manifestation of the ``frustrated'' kinetic term of the Bose-Hubbard model at a negative absolute temperature. We also found the expected N\'eel-type antiferromagnetic ordering in the rhombic limit [cases (g) and (h)]. 
Based on the qualitative agreement between experiments~\cite{negT-exp} and the numerical simulations in the time-dependent GA~\cite{negT-tdGA} for the square lattice, the coherence observed here both in the isotropic and in the rhombic limit implies that the relevant experimental parameter regime can be accessed by the ``negative-T'' approach on the triangular lattice. This observation is the first conclusion of this work.

Curiously, for a certain value of the anisotropy [case (f)], no (quasiclassical) coherence is found up to $600$ ms after the quench. 
This observation may be related to the conjectured spin-liquid behavior of the quantum XY model~\cite{shaking-EPL,scatl,XY-MSWT}.
We were not able to rule out whether the absence of coherence is an artifact of the approximations. Since addressing equilibrium states of the Hamiltonian (\ref{eq:def:BHM}) with frustrated hopping is quite challenging with unbiased numerical methods like QMC or PEPS, at the moment it is not possible to quantitatively validate the results of these numerical simulations. 
The second conclusion is thus that only experiments can verify the phase diagram put forward in Ref.~\cite{shaking-EPL}.

We did not study the chirality of the spiral order in detail, as in a layered setup with independent layers and TOF images taken vertically, this feature cannot be accessed easily.

An interesting future direction would be the generalization of the Feynman relation~\cite{FeynmanStatMech} to unconventional superfluids on the triangular lattice. This relation gives a variational estimation of the dispersion relation of the low-energy excitations $E_{\bq}$ as a ratio of the noninteracting kinetic energy $\epsilon_{\bk}$ and the form factor $S(\bq)$ (Fourier transform of the density-density correlation function). In particular,
Feynman was able to reproduce approximately the phonon-roton spectrum for superfluid He using the form factor measured by neutron scattering.
For ultracold atoms, the form factor could be extracted from noise correlations of TOF images~\cite{Blochreview,noisecorr}, which
could be used to reveal the low-energy dispersion relation of the excitations, in particular, the dynamical exponent $z$ from the relation
$E_{\bq} \sim q^z$ at low momenta.

\textbf{Acknowledgements.} I am grateful for discussions with Hendrik Weimer, Ricardo Doretto, Temo Vekua, Luis Santos, and Ulrich Schneider. I thank especially Stephan Mandt for a critical reading of the manuscript. This research was supported financially by the cluster of excellence QUEST.


\begin{thebibliography}{99}
 \bibitem{Wannier} G. H. Wannier,
 Phys. Rev. \textbf{79}, 357 (1950).

%
 \bibitem{spinice} D. J. P. Morris, D. A. Tennant, S. A. Grigera, B. Klemke, C. Castelnovo, R. Moessner, C. Czter-nasty, M. Meissner, K. C. Rule, J.-U. Hoffmann, K. Kiefer, S. Gerischer, D. Slobinsky, and R. S. Perry 
 Science \textbf{326}, 411 (2009). 

 
  \bibitem{Dirac} P. A. M.  Dirac,
  Proc. Roy. Soc. (London) A \textbf{133}, 60 (1931).
 
  \bibitem{Balents} L. Balents,
  Nature \textbf{464}, 199 (2010).

  
  \bibitem{Blochreview} I. Bloch, J. Dalibard, and W. Zwerger, Rev. Mod. Phys. \textbf{80}, 885 (2008).
  
  
  \bibitem{multibandBloch} T. M\"uller, S. F\"olling, A. Widera, and I. Bloch,
    Phys. Rev. Lett. \textbf{99}, 200405 (2007).
  
  \bibitem{multibandHemmerich} M. \"Olschl\"ager, G. Wirth, and A. Hemmerich,
    Phys. Rev. Lett. \textbf{106}, 015302 (2011).
  
  \bibitem{NeelAFM} A. Koetsier, R. A. Duine, I. Bloch, and H. T. C. Stoof,
    Phys. Rev. A \textbf{77}, 023623 (2008).
    
 \bibitem{shaking0} A. Eckardt, C. Weiss, and M. Holthaus, 
  Phys. Rev. Lett. \textbf{95}, 260404 (2005).
 
 \bibitem{shaking-EPL} A. Eckardt \textit{et al.}, EPL \textbf{89}, 10010 (2010).

  \bibitem{FisherFisher} M. P. A. Fisher, P. B. Weichman, G. Grinstein, and D. S. Fisher,
  Phys. Rev. B \textbf{40}, 546 (1989).
  
 \bibitem{classical-triang-XY} S. Miyashita and H. Shiba,
 Phys. Soc. Jpn. \textbf{53}, 1145 (1984).
 
 \bibitem{quantum-triang-XY} R. Schmied, T. Roscilde, V. Murg, D. Porras, and J. I. Cirac,
 New J. Phys. \textbf{10} 045017, (2008).

 \bibitem{scatl} P. Hauke, Phys. Rev. B \textbf{87}, 014415 (2013). 
 
 \bibitem{shaking-exp} J. Struck, C. \"Olschl\"ager, R. Le Targat, P. Soltan-Panahi, A. Eckardt, M. Lewenstein, P. Windpassinger, and K. Sengstock, Science \textbf{333}, 996 (2011).

 \bibitem{shaking-exp2} J. Struck, M. Weinberg, C. \"Olschl\"ager, P. Windpassinger, J. Simonet, K. Sengstock, R. H\"oppner, P. Hauke, A. Eckardt, M. Lewenstein, and L. Mathey, Nature Physics \textbf{9}, 738 (2013).

\bibitem{Landau} L. D. Landau and E. M. Lifshitz, \textit{Statistical Physics Part 1} (Pergamon, New York, 1980), 3rd ed.    
    
 \bibitem{Mosk} A. P. Mosk, Phys. Rev. Lett. \textbf{95}, 040403 (2005).
 
 \bibitem{negT} \'A. Rapp, S. Mandt, and A. Rosch, Phys. Rev. Lett. \textbf{105}, 220405 (2010).
 
 \bibitem{negT-sim} \'A. Rapp, Phys. Rev. A \textbf{85}, 043612 (2012).
 
 \bibitem{negT-exp} S. Braun, J. P. Ronzheimer, M. Schreiber, S. S. Hodgman, T. Rom, I. Bloch, and U. Schneider, Science \textbf{339}, 52 (2013).
 
 \bibitem{negT-tdGA} \'A. Rapp, Phys. Rev. A \textbf{87}, 043611 (2013).
 
 \bibitem{negT-1D} S. Mandt, A. E. Feiguin, S. R. Manmana, Phys. Rev. A \textbf{88}, 043643 (2013).

 
    
\bibitem{PEPS} F. Verstraete and J. I. Cirac, preprint, cond-mat/0407066 (2004).

\bibitem{QMC} L. Pollet, K. Van Houcke, S. M. A. Rombouts, J. of Comp. Phys. \textbf{225}, 2249 (2007).
 
 \bibitem{triang-NJP} C. Becker \textit{et al.}, New J. Phys. \textbf{12}, 065025 (2010).
 
 \bibitem{fullanisotropy} Deploying full anisotropy in experiments should not present a problem.
 
 \bibitem{asB} The values of the scattering length $a_s$ used here for $^{40}$K can be inverted for the magnetic field value $B$ using the parameters of the corresponding Feshbach resonance.

 \bibitem{verticalbeam} The vertical standing wave should be detuned by a few MHz from the horizontal running beams.
 
 \bibitem{negativeV0U}  Experimentally, the negative scattering length can be reached by ramping the magnetic field through a Feshbach resonance. The anti-trapping potential is provided mainly by the beam profile of the blue-detuned vertical optical lattice beams.
  
 \bibitem{TDG-1} D. Jaksch, V. Venturi, J. I. Cirac, C. J. Williams, and P. Zoller,
	Phys. Rev. Lett. \textbf{89}, 040402 (2002).

 \bibitem{TDG-2} J. Zakrzewski,
	Phys. Rev. A \textbf{71}, 043601 (2005).

 \bibitem{TDG-3} M. Snoek,
	EPL \textbf{95}, 30006 (2011).

 \bibitem{TDG-4} S. S. Natu, K. R. A. Hazzard, and E. J. Mueller,
	Phys. Rev. Lett. \textbf{106}, 125301 (2011). 
 
\bibitem{TDG-5}
    M. Jreissaty, J. Carrasquilla, F. A. Wolf, and M. Rigol,
    Phys. Rev. A \textbf{84}, 043610 (2011).
 
\bibitem{TDG-6}
    J.-S. Bernier, D. Poletti, P. Barmettler, G. Roux, and C. Kollath,
    Phys. Rev. A \textbf{85}, 033641 (2012).


 \bibitem{XY-MSWT} P. Hauke, T. Roscilde, V. Murg, J. I. Cirac, and R. Schmied,
  New J. of Physics, \textbf{12}, 053036 (2010).

  \bibitem{FeynmanStatMech} R. P. Feynman: \textit{Statistical Mechanics}, 2nd ed. (Westview Press, 1998).
  
  \bibitem{noisecorr} E. Altman, E. Demler, and M. D. Lukin,
   Phys. Rev. A \textbf{70}, 013603 (2004).

 
 
\end{thebibliography}
\end{document}